\documentclass[%
 reprint,
superscriptaddress,
 amsmath,amssymb,
 aps,
prb,
floatfix,
nofootinbib
]{revtex4-1}

\usepackage[utf8]{inputenc}
\usepackage{graphicx}
\usepackage{float}
\usepackage{url}

\usepackage{xcolor}
\usepackage{multirow}
\usepackage{xspace}
\usepackage{xfrac}
\usepackage{acro}

\usepackage[OT2,T1]{fontenc}
\usepackage{substitutefont}
\usepackage[inline]{enumitem}

\substitutefont{OT2}{\rmdefault}{wncyr}

\DeclareAcronym{FCM}{short=FCM, long=force constant matrix}
\DeclareAcronym{MLIP}{short=MLIP, long=machine learning interatomic potential}
\DeclareAcronym{NDSC}{short=NDSC, long=non-diagonal supercell}
\DeclareAcronym{DSC}{short=DSC, long=diagonal supercell}
\DeclareAcronym{BZ}{short=BZ, long=Brillouin zone}
\DeclareAcronym{GAP}{short=GAP, long=Gaussian Approximation Potential}
\DeclareAcronym{SOAP}{short=SOAP, long=Smooth Overlap of Atomic Positions}
\DeclareAcronym{DFT}{short=DFT, long=Density Functional Theory}
\DeclareAcronym{RMSE}{short=RMSE, long=root mean squared error}
\DeclareAcronym{PES}{short=PES, long=potential energy surface}

\newcommand{\abinitio}{\textit{ab initio\xspace}}
\newcommand{\etal}{\textit{et al\xspace}}

\begin{document}
\title{Optimal data generation for machine learned interatomic potentials}
\author{Connor Allen}
\affiliation{Department of Physics, University of Warwick, Coventry CV4 7AL, United Kingdom}
\author{Albert P. Bart\'ok}
\affiliation{Department of Physics, University of Warwick, Coventry CV4 7AL, United Kingdom}
\affiliation{Warwick Centre for Predictive Modelling, School of Engineering, University of Warwick, Coventry CV4 7AL, United Kingdom}

\begin{abstract}
\Acp{MLIP} are routinely used atomic simulations, but generating databases of atomic configurations used in fitting these models is a laborious process, requiring significant computational and human effort. 
A computationally efficient method is presented to generate databases of atomic configurations that contain optimal information on the small-displacement regime of the \acl{PES} of bulk crystalline matter.
Utilising \ac{NDSC}\cite{lloyd-williams_lattice_2015}, an automatic process is suggested for \abinitio{} data generation.
\Acp{MLIP} were fitted for Al, W, Mg and Si, which very closely reproduce the \abinitio{} phonon and elastic properties.
The protocol can be easily adapted to other materials and can be inserted in the workflow of any flavour of \ac{MLIP} generation. \\

\centering{
{\small
UK Ministry of Defence © British Crown Owned Copyright 2022/AWE
}}

\end{abstract}

\maketitle

\section{Introduction}
Modern approaches to material discovery and characterisation include the use of \abinitio{} modelling.
While well-established methods, such as \ac{DFT}, reliably predict the electronic, mechanic and thermodynamic properties of materials\cite{Pickard.2006,Pickard.2011}, most of these techniques are limited by the fact that computational effort scales as $\mathcal{O}(N^3)$ or worse with the number of atoms ($N$).
Although linear scaling implementations of \ac{DFT} exist\cite{ordejon_linear_1995,Prentice.2020c7k}, large prefactors prevent efficient sampling of atomic configurations, which are required, for example, to compute thermodynamic averages. 
In the past decade, data driven approaches emerged as possible solutions to realise \abinitio{} accuracy at an affordable computational cost, even at large length and time scales\cite{Deringer.2020,Cheng.2019}.
Surrogate models of the Born-Oppenheimer \acf{PES} can be generated in the form of \acp{MLIP}\cite{blank_neural_1995, bartok_gaussian_2010, seko_first-principles_2015,Deringer.2021}.
These are based on non-linear, non-parametric regression of the PES, fitted using databases of atomic configuration and their associated \abinitio{} total energies and derivatives.
Exploiting locality, or the nearsightedness of quantum mechanics\cite{Prodan.2005}, fitting can be performed on configurations containing relatively few atoms, therefore keeping the computational cost of generating the database affordable, while the resulting \ac{MLIP} may be used in extended systems.
Machine learning techniques in atomic modelling have evolved into a mature field, with a broad range of methods present, such as Schnet\cite{Schuett.2018}, MTP\cite{Shapeev.2016}, ACE\cite{Drautz.2019}, NN\cite{Behler.2017}, PhysNet\cite{Unke.2019} and \ac{GAP}\cite{Bartok.2015}, among others.
While the underlying principles of \acp{MLIP} can vary significantly, they all rely on carefully built databases that contain atomic configurations representative of a wide range of atomic environment that are relevant to the intended purpose of the model.

Creating such databases of atomic configurations are time consuming, both in terms of human and computational effort.
Even though automated approaches, such as active learning\cite{Deringer.2017e5c,Deringer.202038} can eliminate human intervention to a large extent, ``hand-crafting'' parts of the database is often necessary to include specific configurations, such as various known crystalline polymorphs, defects or surfaces.
Accurate modelling of the elastic and vibrational properties of bulk crystals is crucial in numerous applications, such as the finite temperature stability of different phases or defect formation energies.
To provide targeted fitting data for the phonon spectrum, samples from molecular dynamics calculations\cite{Bartok.2018} or specifically perturbed configurations\cite{george_combining_2020} are employed routinely.
In this work, we suggest a highly efficient approach based on \acp{NDSC} introduced by Lloyd-Williams and Monserrat\cite{lloyd-williams_lattice_2015}, which can be used to automatically generate small atomic configurations that contain optimal information to fit the PES in the small-displacment regime.
As \abinitio{} calculations only need to be performed on configurations containing only a handful of atoms each, data generation is efficient. We used the \ac{GAP} framework\cite{Bartok.2010} to fit \acp{MLIP} of bulk crystals of metallic and semiconducting elements representing different crystal structures.
In our benchmarks, we obtained highly accurate phonon dispersions and elastic properties when comparing to the underlying \ac{DFT} model.

\section{Background}

\subsection{\acl{GAP}}\label{sec:gap}
The machine learned potential framework we use is \ac{GAP}\cite{bartok_gaussian_2010}, although we emphasise that the database generation workflow is easily transferable to other approaches.
\Ac{GAP} can be formulated as a kernel based method that predicts the total energy of a given configuration $\mathbf{X} = \{ \mathcal{R}_1, \mathcal{R}_2 ,... \mathcal{R}_I\}$ as:
\begin{equation}\label{eq:total_energy}
    E(\mathbf{X}) =  \sum_{i}^{I}\sum_{s}^{M}\alpha_s K(\mathcal{R}_i,\mathcal{R}_s) 
\end{equation}
where $\mathcal{R}$ represents an atomic environment, $s$ is a summation over a set of $M$ representative environments, each associated with a weight $\alpha_s$.
The kernel function, $K(\mathcal{R},\mathcal{R}')$, may be regarded as a similarity measure between two atomic environments $\mathcal{R}$ and $\mathcal{R}'$.
In this work, we describe atomic environments using the \ac{SOAP}\cite{bartok_representing_2013,Musil.2021} descriptor, where a given atomic neighbourhood environment is initially characterised as a density:
\begin{equation}\label{eq:neighbour_density}
    \rho_i(\mathbf{r}) = \sum_{i'} f_{\textrm{cut}}(r_{ii'})e^{-|\mathbf{r}-\mathbf{r}_{ii'}|^{2}/2\sigma_{\textrm{atom}}^2}
\end{equation}
where a Gaussian with a width of $\sigma_{\textrm{atom}}$ is centered on each atom up to a specified cutoff radius, whereby beyond this cutoff, $f_{\textrm{cut}}$ smoothly goes to zero.
This density is then expressed in terms of radial and spherical harmonics basis functions
\begin{equation*}
    \rho_i(\mathbf{r}) = \sum_{nlm} c_{nlm}^i Y_{lm}(\hat{\mathbf{r}})g_n(\mathbf{r})
\end{equation*}
which are defined up to a specified complexity controlled by $n_{max}$ and $l_{max}$ and $m=-l, -l+1, ... \ l$.
Rotationally invariant features are constructed from the power spectrum elements as
\begin{equation*}
   \tilde{\mathbf{p}}_i \equiv \sum_{m=-l}^l c_{nlm}^{i*} c_{n'lm}^i \\
\end{equation*}
which are normalised
\begin{equation*}
    \mathbf{p}_i = \tilde{\mathbf{p}}_i / |\tilde{\mathbf{p}}_i |
    \textrm{.}
\end{equation*}
Finally, we obtain an expression for our covariance evaluation between atomic neighbourhoods as
\begin{equation}\label{eq:kernel}
    K(\mathcal{R}, \mathcal{R}') = \delta^2(\mathbf{p} \cdot \mathbf{p}')^\zeta
\end{equation}
where $\delta$ and $\zeta$ are hyperparameters that control the energy scaling of the descriptor
and smoothness of the kernel, respectively.

To obtain the weights $\alpha_s$, we minimise the loss function
\begin{equation}\label{eq:loss_func}
    \mathcal{L}= \sum_{n=1}^{N}\frac{[y_n - \tilde{y}_n]^2}{\sigma_n^2} + \sum_{s,s'}^{M} \alpha_{s} K(\mathcal{R}_s, \mathcal{R}_{s'})\alpha_{s'}
\end{equation}
where the quantity $y$ can be the one of total energy, force or stress value of an atomic configuration, and $\sigma_n$ is a hyperparameter, related to the weight or importance of each data point.
$y_n$ represents the reference \abinitio{} values, whereas $\tilde{y}_n$ is the \ac{GAP} prediction of the total energy using equation~\ref{eq:total_energy} or the appropriate derivatives, with respect to atomic coordinates or lattice deformations.
This definition allows us to fit using the total energy observations, as well as the forces on atoms and the virial stress for each configuration. 
The second term in the loss function acts as a regulariser.

In algebraic form, the minimisation of this loss function with respect to $\alpha$ yields
\begin{equation*}
    \mathbf{\alpha} = (\mathbf{K}_{MM} + \mathbf{K}_{MN} \mathbf{\Sigma}^{-1} \mathbf{K}_{NM})^{-1} \mathbf{K}_{MN} \mathbf{\Sigma}^{-1}  \mathbf{y} 
\end{equation*}
where $\boldsymbol{\Sigma}$ contains a diagonal matrix containing the values of $\sigma_n$ and $\mathbf{y} = [y_1, \ldots, y_N]$.
The kernel matrices $\mathbf{K}$ contain all pairwise evaluations of kernel functions between atomic environments, where $M$ denotes the representative set, and $N$ refers to the reference database of atomic configurations.

\subsection{Non-diagonal supercells}
For an interatomic potential model to represent the \ac{PES} near a stationary point, which in our case is the perfect bulk crystal, it needs to reproduce the \ac{FCM} of an extended system, formulated as the Hessian of the Born-Oppenheimer total energy $E$ with respect to Cartesian atomic coordinates
\begin{equation*}
    \Phi_{i\alpha j \beta} = \frac{\partial^2 E}{\partial r_{i\alpha} \partial r_{j\beta}}
\end{equation*}
where $i$, $j$ denote atomic indices, and $\alpha$, $\beta$ represent Cartesian directions.
Under the harmonic approximation, the total energy is expressed as a Taylor expansion with terms higher than second order truncated.
Most \ac{MLIP} approaches rely on the assumption of locality of the atomic interactions, i.e. for any small number $\varepsilon >0$ there exist an $r_\textrm{cut}$ such that all $\Phi_{i\alpha j \beta} < \varepsilon$ for $|\mathbf{r}_i-\mathbf{r}_j| > r_\textrm{cut}$, corresponding to a truncated \ac{FCM}.
It should be noted that \ac{MLIP} frameworks can be extended to represent long-range, such as Coulombic, interactions, but our current discussion is limited to the short-range term representing local, i.e. covalent or metallic, bonding.

It is customary to express the elements of the \ac{FCM} such that they are indexed by labels of the basis atoms $i$ and $j$ within their primitive unit cells, and the displacement vector $\mathbf{R}_p$ that translates the two primitive unit cells into each other:
\begin{equation*}
     \Phi_{i\alpha j \beta}(\mathbf{R}_p) \equiv \Phi_{i\alpha j' \beta} 
\end{equation*}
such that $\mathbf{r}_{j'} = \mathbf{R}_p + \mathbf{r}_j$.
Fitting \acp{MLIP} is ultimately data driven, therefore atomic configurations should ideally contain information on as many elements of the truncated \ac{FCM} as possible.
As fitting data is most commonly provided as atomic configurations with the corresponding \abinitio total energies, forces, and stresses, supercells capable of accommodating perturbations of distant atom pairs are highly desirable.
A common approach is to use supercells generated such that their shape is as closely cubic as possible, as an attempt to include atom pairs isotropically.
The lattice vectors $\mathbf{a}_s$, $\mathbf{b}_s$, and $\mathbf{c}_s$ of a supercell $s$ are related to the unit cell lattice vectors $\mathbf{a}_u$, $\mathbf{b}_u$, and $\mathbf{c}_u$ as
\begin{equation*}
\begin{pmatrix}
\boldsymbol{a}_s \\ \boldsymbol{b}_s \\ \boldsymbol{c}_s
\end{pmatrix}=
\begin{pmatrix}
S_{11} & S_{12} & S_{13} \\
S_{21} & S_{22} & S_{23} \\
S_{31} & S_{32} & S_{33}
\end{pmatrix}
\begin{pmatrix}
\boldsymbol{a}_u \\ \boldsymbol{b}_u \\ \boldsymbol{c}_u
\end{pmatrix}
\end{equation*}
\noindent where elements of the supercell matrix $S_{ij} \in \mathbb{Z} $, and for a \ac{DSC} $S_{ij} = S_i \delta_{ij}$.
For example, generating \ac{DSC} of the cubic unit cell of fcc or bcc crystals, or in case of hexagonal crystals, using the orthorhombic unit cell is a convenient choice, as the unit cell lattice vectors are orthogonal.

Atoms need to be displaced in the supercell before computing the \abinitio{} total energy and its derivatives.
George \etal{}\cite{george_combining_2020} suggested using randomly perturbed atomic coordinates as well as the displacement of a single atom in the supercell.
Randomising displacements with a certain amplitude is expected to result in force observations that are dominated by terms containing the largest elements of the \ac{FCM}, as the $\alpha$ component of force of atom $i$ may be approximated as
\begin{equation}\label{eq:harmonicforce}
    f_{i\alpha} \approx -\sum_{j\beta} \Phi_{i\alpha j \beta} (r_{j\beta} - r_{j\beta,0})
\end{equation}
where $\mathbf{r}_{j,0}$ denotes the equilibrium position of atom $j$ in the supercell.
Since fitting of \acp{MLIP} assumes some degree of uncertainty on each observation as described in Section~\ref{sec:gap}, such dominance may have detrimental effect on the quality of the fit as small contributions will be indistinguishable from noise.
More terms in equation~\ref{eq:harmonicforce}, corresponding to larger supercells, is expected to aggravate the situation, leading to poor fit of small elements of the \ac{FCM}.
Alternatively, the displacement of a single atom along a Cartesian direction results in the resolution of each individual element of the \ac{FCM}, but such configurations contain highly correlated atomic environments and cannot be regarded as realistic examples of configurations sampled from finite temperature simulations.
Samples from finite temperature simulations, such as molecular dynamics, are an optimal solution, but only if the sampling uses a sufficiently similar \ac{PES} to that of the \abinitio{} model, otherwise the configurations will be practically equivalent to those generated by randomisation.
The computational cost of the \abinitio{} reference calculations when using \ac{DSC} scales as $(S_1 S_2 S_3)^3$ if using plane-wave \ac{DFT}, therefore the size of the supercell, and the representable elements of the \ac{FCM} is severely limited.

Lloyd-Williams and Monserrat \cite{lloyd-williams_lattice_2015} demonstrated that perturbations that require a \ac{DSC} constituting $S_{1} \times S_{2} \times S_{3} $ primitive cells, may be represented by a \ac{NDSC} with no more than the least common multiple of $S_{1}$, $S_{2}$, and $S_{3}$ number of primitive cells.
Lloyd-Williams and Monserrat suggested this method to sample the vibrational modes in the \ac{BZ} of a crystal uniformly on an $N \times N \times N$ grid.
When computing the \ac{FCM} using finite differences, \ac{DSC} of the size $N \times N \times N$ are needed, whereas if using \acp{NDSC}, only supercells of size up to $N$ are required.
Even though more \acp{NDSC} have to be typically considered, each individual calculation incurs significantly less computational cost, while the process can benefit from trivial parallelisation.
Overall, significant reductions in the the computational cost associated with \abinitio{} phonon dispersion calculations can be realised, and also allows one to consider more dense sampling of the \ac{BZ}.

\subsection{Phonon dispersion}
With the force constants determined under the harmonic approximation, one method for finding the frequency of the allowed vibrational modes $\mathbf{q} \in \textrm{\ac{BZ}}$ is done via finding the eigenvalues of the dynamical matrix $\mathbf{D}(\mathbf{q})$ whose elements are obtained via Fourier-transforming the mass-weighted \ac{FCM} as
\begin{equation}\label{dynamical_matrix}
D_{i\alpha j \beta} (\mathbf{q}) = \frac{1}{\sqrt{m_{i}m_{j}}} \sum_{\textbf{R}_{p}} \Phi_{i\alpha j \beta}(\mathbf{R}_p) e^{-i \mathbf{q} \cdot \mathbf{R}_p}
\end{equation}
where $m_i$ and $m_j$ are the masses of atoms $i$ and $j$.
The square root of the eigenvalues at each $\mathbf{q}$ vector are the phonon frequencies.
Negative eigenvalues result in imaginary frequencies, corresponding to dynamically unstable modes, along which displacements result in lowering the energy.
As customary, we represent such imaginary frequencies as negative numbers on our phonon dispersion plots.

\section{Methodology}
\subsection{Density Functional Theory calculations}
The underlying \abinitio{} calculations that were used to train the interatomic potentials as well as benchmark them was preformed using the plane-wave \ac{DFT} code, \verb_CASTEP_ \cite{clark_first_2005}.
On-the-fly ultrasoft pseudopotentials \cite{vanderbilt_soft_1990} were generated for Mg, Al, Si, and W with the respective valence electronic structure: 2s$^2$2p$^6$3s$^2$,  3s$^2$3p$^1$, 3s$^2$3p$^2$, and 5s$^2$5p$^6$4f$^{14}$6s$^2$5d$^4$.
In all instances a generalized gradient approximation \cite{perdew_generalized_1996} exchange-correlation functional was used.
The plane-wave energy cutoff ($\textrm{E}_{\textrm{cut}}$), density of the electronic \ac{BZ} sampling of a Monkhorst-Pack grid \cite{monkhorst_special_1976} ($k$-spacing) and the self-consistent field energy tolerance for convergence was set for each system to find a converged result on the total energy and derivative quantities.
Geometry optimisations were then performed for all systems to find the relaxed lattice parameters for a given fixed crystal symmetry. The specific DFT parameter set and primitive cell information found from the geometry optimisations are presented in Table~\ref{tab:dft_paramset}. 

\begin{table}[H]
    \centering
    \begin{tabular}{||l||c|c|c|c||}
         \hline
         & Mg & Al   & Si  & W \\
         \hline
         \footnotesize{\textbf{DFT:}} & & & &  \\
         \ $E_\textrm{cut}$ (eV) & 520 & 800   & 400 & 600\\
         \ $k$-spacing (\AA$^{-1}$) & 0.012 & 0.010   & 0.030 & 0.015 \\
         \ SCF tol. (eV)  &$10^{-11}$ & $10^{-11}$  &$10^{-11}$ & $10^{-10}$   \\
         \footnotesize{\textbf{Lattice:}} & & & &\\
         \ Structure & hcp & fcc & dia. & bcc\\
         \ $a$ (\AA) & 3.198 & 2.856 & 3.867 & 2.756\\
         \ $c$ (\AA) & 5.179 & - & - & -\\
         \hline
    \end{tabular}
    \caption{\Ac{DFT} parameter set (planewave cutoff energy, spacing of the k-point sampling of the \ac{BZ} and tolerance of the self-consistent iterations) used to perform ab initio data generation and benchmark comparison and the geometry optimised lattice information for each system. }
    \label{tab:dft_paramset}
\end{table}

\noindent The elements of the \ac{FCM} for the phonon dispersion calculations at the \abinitio{} level were determined  using the finite difference 
method \cite{kunc_ab_1982} as implemented in  \verb_CASTEP_, corresponding to a $4\times 4 \times 4$ grid in the \ac{BZ}.
A  displacement of 0.05 \AA \ from the ideal lattice site was used, and phonon dispersion curves were computed along high symmetry lines using Fourier interpolation.

\subsection{Database generation}\label{sec:database}
Our aim is to investigate a protocol that produces database configurations targeted to fit vibrational properties of crystalline materials in a computationally optimal way.
We suggest basing the workflow on \acp{NDSC}, which can represent long-range perturbation of crystalline order using the configurations that contain the fewest possible atoms.

To construct a database that can explores displacements around the pristine crystal geometry corresponding to an $N\times N \times N$ supercell of the primitive unit cell, we generated \acp{NDSC} using the \verb_FORTRAN 90_ program by Lloyd-Williams and Monserrat \cite{lloyd-williams_lattice_2015} which contain supercells formed of up to $N$ primitive unit cells.
The \ac{NDSC} configurations therefore contain information about the  vibrational modes corresponding to a $N\times N \times N$ phonon $\mathbf{q}$-vector grid.
In addition, we introduced deformation of the cells by homogeneous scaling of the cell vectors to capture isotropic compression and expansion.
To capture the response of atoms displaced from ideal lattice sites within the different \ac{NDSC} configurations, copies were made where atoms were randomly displaced via a normal distribution with standard deviation of 0.10 $\textrm{\AA}$.
Finally, to inform the fitting procedure on how the \ac{PES} responds to anisotropic cell deformations, random shearing was applied on the \ac{NDSC} configurations.
The lattice vectors, contained in $\mathbf{L}$ were transformed by a symmetrical strain matrix, $\boldsymbol{\epsilon}$, as:
\begin{equation*}
    \mathbf{L}_{\textrm{rand.}} = (\mathbf{I} + \boldsymbol{\epsilon})\mathbf{L} 
\end{equation*}
where $\mathbf{I}$ is the identity matrix and each entry of the strain matrix is sample from a uniform distribution, $\epsilon_{ij}\sim \mathcal{U}(-0.01, 0.01)$, such that $\boldsymbol{\epsilon}$ is symmetric.

To investigate the transferability of the proposed workflow for database generation using \ac{NDSC}, four different crystal structures were considered: hexagonal close-packed (hcp) Mg, diamond (dia) Si, body-centred cubic (bcc) W and face-centred cubic (fcc) Al.
The \acp{NDSC} were generated were commensurate with a $4 \times 4 \times 4$ grid sampling of the vibrational \ac{BZ} of the relaxed primitive cell of each system.
All configuration manipulations were done through the Atomic Simulation Environment\cite{atomic_simulation_environment}.

\subsection{Fitting \acp{MLIP}}\label{sec:fitting}
We used the \ac{GAP} framework to generate \acp{MLIP}, but we stress that any other similar fitting approaches would benefit equally.
For all models presented here we select 1400 sparse points through a CUR decomposition and set $\delta=2\,\textrm{eV}$ and $\zeta=4$. Further \ac{GAP} hyperparameters and details on the training data set are specified in Table~\ref{tab:gap_paramset}. For the Al Bain path model developed, additional data was included to capture the bcc phase and the half-way point on the Bain path as described by equation \ref{eq:bainpath_mat}. This \ac{GAP} was trained on 3751 atomic environments, for a total of 1455 target energies, using 1400 sparse points. For the minimal data case on fcc Al, 74 atomic environments (24 target energies) constituted the training set for the \ac{NDSC} model, whereas 65 atomic environments (2 target energies) were considered for the \ac{DSC} model. In the minimal data GAP for fcc Al, both the NDSC and DSC contained the geometry optimised primitive cell.



\begin{table}[htb]
    \centering
    \begin{tabular}{||l||c|c|c|c||}
         \hline
         & Mg & Al   & Si  & W \\
         \hline
         \footnotesize{\textbf{\ac{GAP}:}} & & & &  \\
         \ $r_\textrm{cut}$ (\AA)  & 8.0  & 10.0 & 6.0 & 6.0\\
         \ $n_\textrm{max}$     & 8    & 10   &  8  & 8  \\
         \ $l_\textrm{max}$     & 6    & 8   &  6  & 6 \\
         \ $\sigma_{\textrm{atom}}$ (\AA) & 0.5 & 0.5 & 0.5 & 0.5 \\
         \footnotesize{\textbf{Regularisation:}} & & & &  \\
         \ $\Delta_{\textrm{F}}$ &  $10^{-2}$ & $10^{-2}$& $10^{-2}$& $10^{-2}$\\
         \ $\sigma_{\textrm{F}}^{\textrm{min}}$ (eV/\AA)&$10^{-3}$ &$10^{-3}$ &$10^{-3}$ &$10^{-3}$\\
         \ $\Delta_{\textrm{V}}$ & $10^{-2}$ & $5\times10^{-3}$& $10^{-2}$& $10^{-2}$\\
         \ $\sigma_{\textrm{V}}^{\textrm{min}}$ (eV) &$10^{-3}$ &$5\times10^{-4}$ &$10^{-3}$ &$10^{-3}$\\
         \footnotesize{\textbf{Amount of Data:}} & & & & \\
         \ $N_{\textrm{atoms}}$ & 4276 & 1683 & 3300 & 1550 \\
         \ $N_{\textrm{energy}}$ & 1004 & 850 & 936  & 836 \\
        \footnotesize{\textbf{Lattice}} & & & & \\
         \ Structure & hcp & fcc & dia. & bcc\\
         \ $a$ (\AA) & 3.194  & 2.856 & 3.867 & 2.757\\
         \ $c$ (\AA) & 5.181  & - & - & -\\
         \hline
    \end{tabular}
    \caption{\ac{GAP} hyperparameter set for each system, and associated data used for training. Virial stresses ($N_{\textrm{virial}} = 6N_{\textrm{energy}}$) and atomic forces ($N_{\textrm{force}} = 3N_{\textrm{atoms}}$) was also included on all configurations. Primitive cell vectors from a geometry optimisation using each potential are also presented.}
    \label{tab:gap_paramset}
\end{table}
Based on the work of George \etal{}\cite{george_combining_2020}, we employed adaptive regularisation, via adjusting the hyperparameter $\sigma$ as described by the loss function in Equation \ref{eq:loss_func}. In addition to scaling $\sigma$ corresponding to force components of the data, we implemented a similar adjustment algorithm  for virial stress components. Throughout this work we use a constant regularisation on the total energy predictions while the element wise viral regularisation and component wise force regularisation are implemented as

\begin{table}[H]
    \centering
    \begin{tabular}{||l|ccc||}
    \hline
    {} & \multicolumn{2}{c}{$\sigma_n$} &{}\\ 
    \hline
    energy & \multicolumn{2}{c}{0.001~eV}&{}\\
    \hline
    \multirow{2}{*}{force} & $\Delta_{F} |\mathbf{F}_i|$ & if  $\Delta_{F} |\mathbf{F}_i| > \sigma_F^{\textrm{min}}$ &{}\\
    & $\sigma_{F}^{\textrm{min}}$ & else &{}\\
    \hline
    \multirow{2}{*}{virial} & $\Delta_{V} |V_{\alpha\beta}|$ & if  $\Delta_{V} |V_{\alpha\beta}| > \sigma_{V}^{\textrm{min}}$ &{}\\
    & $\sigma_{V}^{\textrm{min}}$ & else&{} \\
    \hline
    \end{tabular}
\end{table}
\noindent introducing $\sigma^{\textrm{min}}$ to define a minimum value for the regularisation and $\Delta$ to scale the value for each component of the corresponding quantity. The choice of regularisation parameters are summarised in Table~\ref{tab:gap_paramset}. 
 
The \ac{FCM} elements of the developed \acp{MLIP} were calculated using the finite difference method \cite{kunc_ab_1982} using the \verb_phonopy_ package\cite{togo_first_2015}.
We calculated phonon frequencies along the high symmetry lines suggested by Setyavan and Curtarolo\cite{Setyawan.2010}, and determined the phonon density of states based on a $40\times 40\times 40$ $\mathbf{q}$-vector grid.




\section{Results}

\begin{table}[htb]
    \centering
    \begin{tabular}{||l|c|c|c|c|c|c|c||}
    \hline
    E.C & B & C$_{11}$ & C$_{12}$ & C$_{13}$ & C$_{33}$ & C$_{44}$ & C$_{66}$ \\
    \hline
    \textbf{Mg} & & & & & & &\\
    \ \footnotesize{GAP} & \footnotesize{36.2} &  \footnotesize{65.6} &  \footnotesize{20.7} &  \footnotesize{22.4} &  \footnotesize{63.6} &  \footnotesize{17.4} &  \footnotesize{22.1} \\
    \ \footnotesize{DFT} & \footnotesize{36.5} &  \footnotesize{64.2} &  \footnotesize{20.2} &  \footnotesize{23.7} &  \footnotesize{64.7} &  \footnotesize{17.1} &  \footnotesize{20.9}\\
    \ \footnotesize{Expr.\cite{slutsky_elastic_1957}} & \footnotesize{36.9} &  \footnotesize{63.5} &  \footnotesize{25.9} &  \footnotesize{21.7} &  \footnotesize{66.5} &  \footnotesize{18.4} &  \footnotesize{18.8}\\
\hline
    \textbf{Al} & & & & & & &\\
    \ \footnotesize{GAP} & \footnotesize{77.5} &  \footnotesize{109.2} &  \footnotesize{61.7} &  \footnotesize{-} &  \footnotesize{-} &  \footnotesize{31.9} &  \footnotesize{-}\\
    \ \footnotesize{DFT} & \footnotesize{77.6} &  \footnotesize{106.9} &  \footnotesize{62.9} &  \footnotesize{-} &  \footnotesize{-} &  \footnotesize{33.3} &  \footnotesize{-}\\
    \ \footnotesize{Expr. \cite{vallin_elastic_1964}} & \footnotesize{82.0} &  \footnotesize{116.3} &  \footnotesize{64.8} &  \footnotesize{-} &  \footnotesize{-} &  \footnotesize{30.9} &  \footnotesize{-}\\
\hline
    \textbf{Si} & & & & & & &\\
    \ \footnotesize{GAP} & \footnotesize{88.6} &  \footnotesize{152.4} &  \footnotesize{56.7} &  \footnotesize{-} &  \footnotesize{-} &  \footnotesize{72.4} &  \footnotesize{-}\\
    \ \footnotesize{GAP$^{\textrm{*}}$} & \footnotesize{89.7} &  \footnotesize{155.8} &  \footnotesize{56.7} &  \footnotesize{-} &  \footnotesize{-} &  \footnotesize{72.8} &  \footnotesize{-}\\
    
    \ \footnotesize{DFT} & \footnotesize{88.6} &  \footnotesize{152.7} &  \footnotesize{56.6} &  \footnotesize{-} &  \footnotesize{-} &  \footnotesize{73.3} &  \footnotesize{-}\\
    \ \footnotesize{Expr. \cite{hall_electronic_1967}} & \footnotesize{99.1} &  \footnotesize{167.5} &  \footnotesize{64.9} &  \footnotesize{-} &  \footnotesize{-} &  \footnotesize{80.2} &  \footnotesize{-}\\
\hline
    \textbf{W} & & & & & & &\\
    \ \footnotesize{GAP} & \footnotesize{306.4} &  \footnotesize{510.5} &  \footnotesize{204.3} &  \footnotesize{-} &  \footnotesize{-} &  \footnotesize{136.9} &  \footnotesize{-}\\
    \ \footnotesize{DFT} & \footnotesize{306.3} &  \footnotesize{512.8} &  \footnotesize{203.0} &  \footnotesize{-} &  \footnotesize{-} &  \footnotesize{135.9} &  \footnotesize{-}\\
    \ \footnotesize{Expr. \cite{stathis_elastic_1980}} & \footnotesize{314.7} &  \footnotesize{533.9} &  \footnotesize{205.1} &  \footnotesize{-} &  \footnotesize{-} &  \footnotesize{163.3} &  \footnotesize{-}\\
    \hline
    \end{tabular}
    \caption{Comparison of elastic constant predictions for \ac{NDSC} \ac{GAP} developed for each system, the underlying \ac{DFT} calculations and experiment. GAP$^*$ refers to the reduced model in Si that does not contain \ac{NDSC} configurations.}
    \label{tab:geom_elastic}
\end{table}

Having fitted a series of \ac{GAP} models for W, Al, Si and Mg using databases consisting of \ac{NDSC} configurations, we evaluated the accuracy of each model by comparing its vibrational and elastic properties to \ac{DFT} values.
Our reference \ac{DFT} calculations show good agreement with the literature\cite{debernardi_ab_2001,szlachta_accuracy_2014,jiang_structural_2020,togo_first_2015,zhuang_elastic_2016,george_combining_2020}.
Overall, we find that all fitted models show excellent performance in our benchmarks.
The summary of geometric and elastic parameters predicted by the \ac{GAP} models, and comparisons to \ac{DFT} results is presented in Table~\ref{tab:geom_elastic}.
Excellent agreement with \ac{DFT} may be observed across all our test systems, with the \ac{RMSE} on phonon modes below 0.5 THz. 

\begin{figure}[htb]
    \centering
    \includegraphics[width=\linewidth]{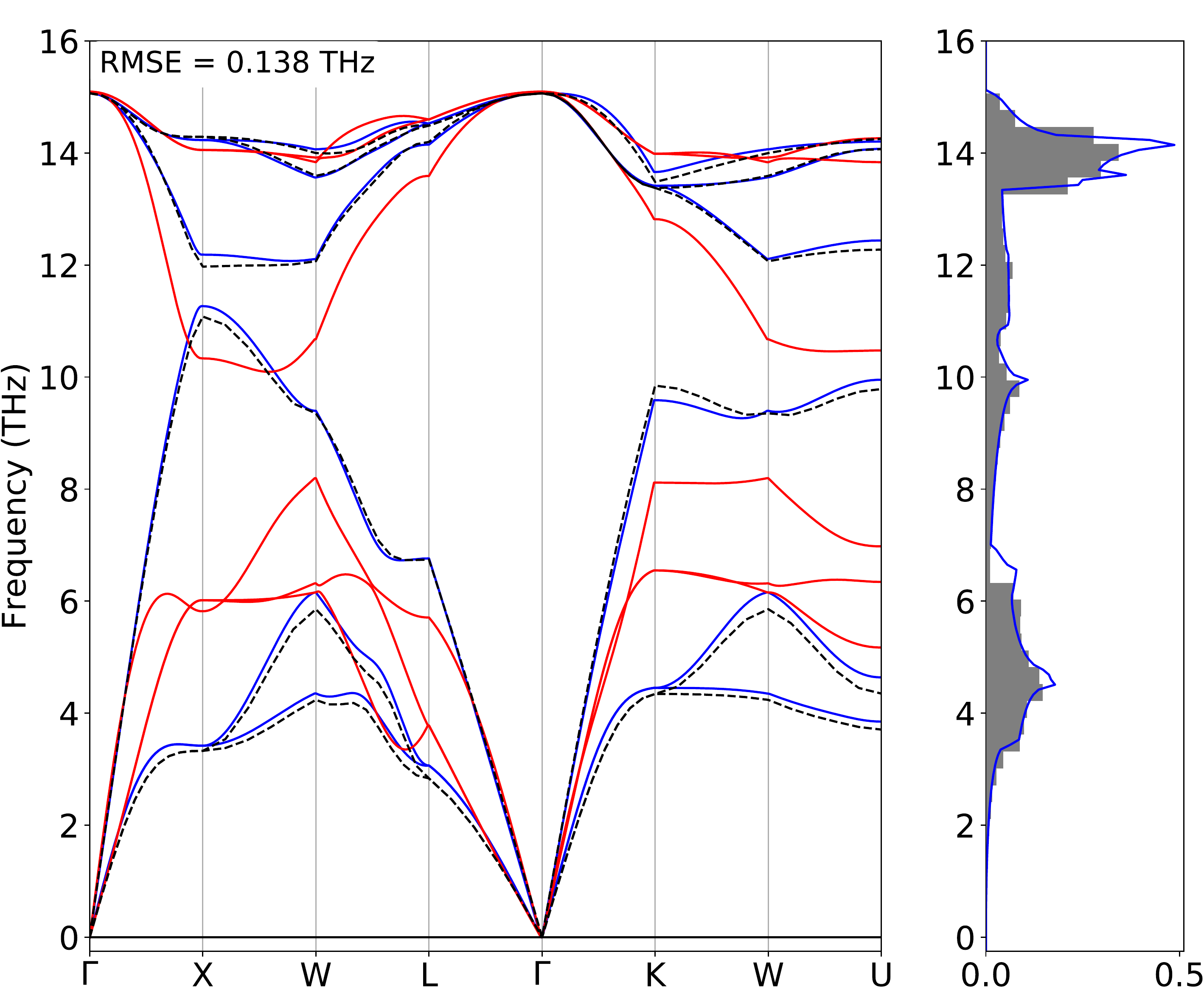}
    \caption{Phonon dispersion and density of states calculations by \acp{GAP} trained using a minimal database only containing deformed primitive unit cells (red) and the full database containing \ac{NDSC} configurations (blue). Reference \ac{DFT} values are shown as black dashed lines.}
    \label{fig:Si_dia_phonons}
\end{figure}

We fitted a reduced model for Si that only contained the primitive unit cell configurations, in order to study the role of different elements of the database. Tabulated results in Table~\ref{tab:geom_elastic} show excellent agreement of elastic constants for both models.
As the elastic moduli are related to the slope of the acoustic phonon modes near the $\Gamma$-point\cite{Boer.2018}, portions of the dispersion of phonon modes are also in good agreement for the minimal model, as shown in Figure~\ref{fig:Si_dia_phonons}.
However, at phonon modes corresponding to intermediate wavelengths the agreement for the minimal model is poor, confirming that deformed unit cells provide information to the \ac{GAP} fitting about the elastic behaviour of a given material, but larger supercells are required to inform the fitting procedure on the full \ac{FCM}.
Indeed, adding \ac{NDSC} configurations to the database, we recover the phonon dispersion across the \ac{BZ} accurately.


\begin{figure}[h!tb]
    \centering
     \includegraphics[width=\linewidth]{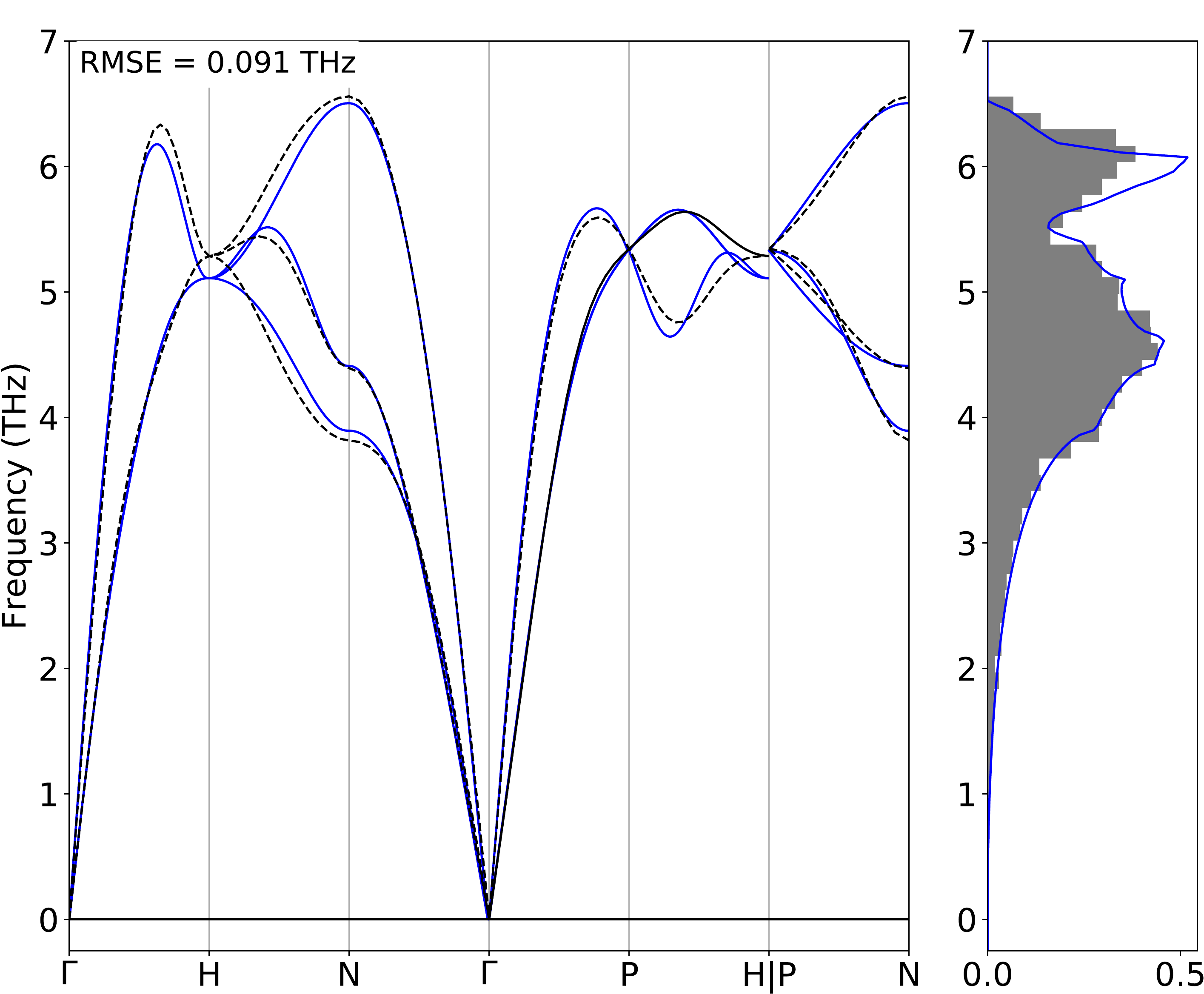}
     \includegraphics[width=\linewidth]{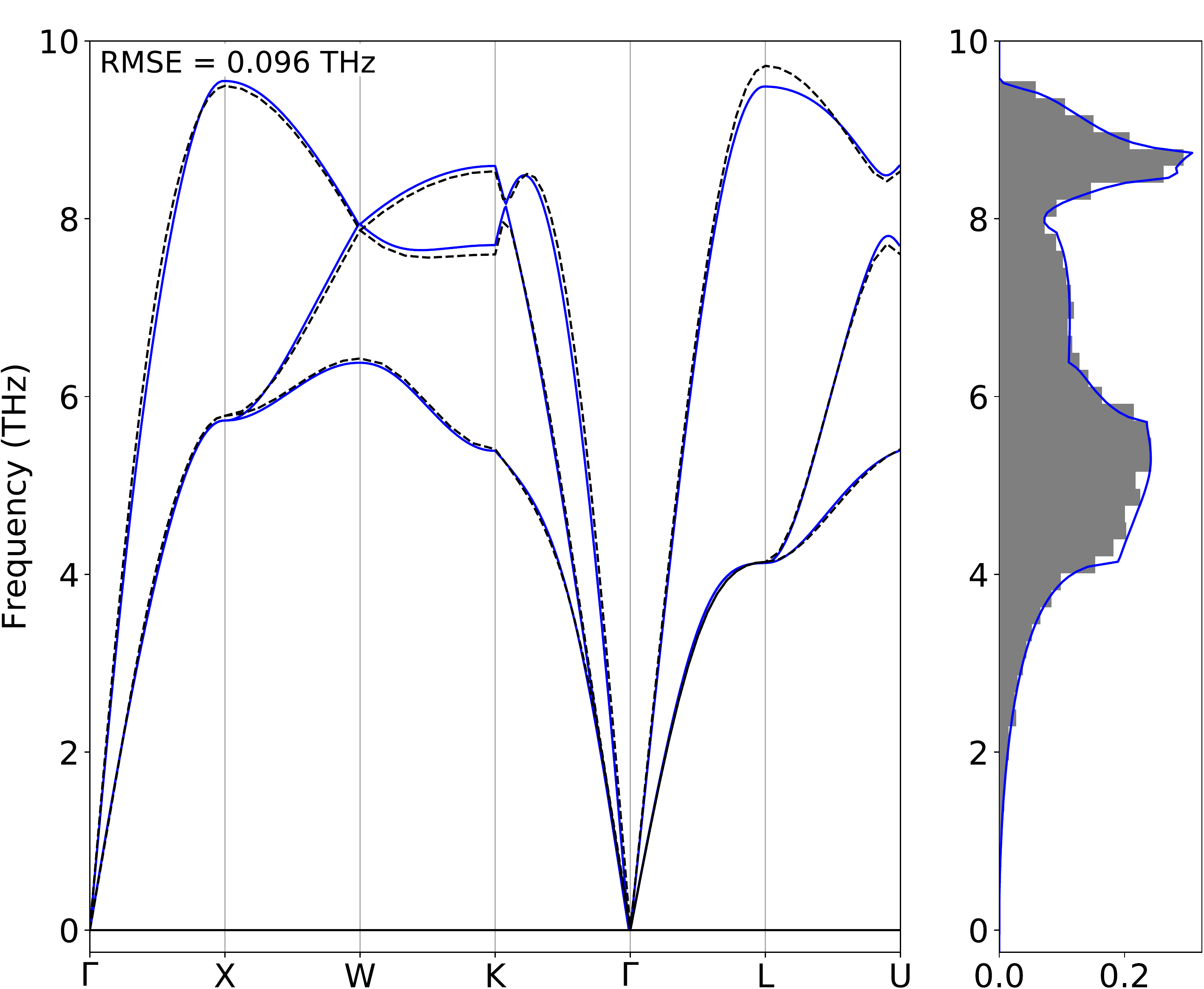}
    \caption{\ac{GAP} models trained using \ac{NDSC} structures (blue) versus commensurate \ac{DFT} (black dashed) phonon dispersion calculations for bcc W (top panel) and fcc Al (bottom panel) with corresponding density of states.}
    \label{fig:W_bcc_phonons}
\end{figure}

The \ac{GAP} model reproduces the elastic and vibrational properties of bcc W to a great accuracy, although we note that the largest error compared to \ac{DFT} in the phonon frequencies is at the $H$ point, together with a discrepancy in the curvature  along the $P-H$ direction.
Since the phonon mode at $H$ corresponds to displacing oppositely the two atoms located on neighbouring corners of the cubic cell, the corresponding elements of the \ac{FCM} can be regarded as well represented in our database.

\begin{figure}[htb]
    \centering
    \includegraphics[width=\linewidth]{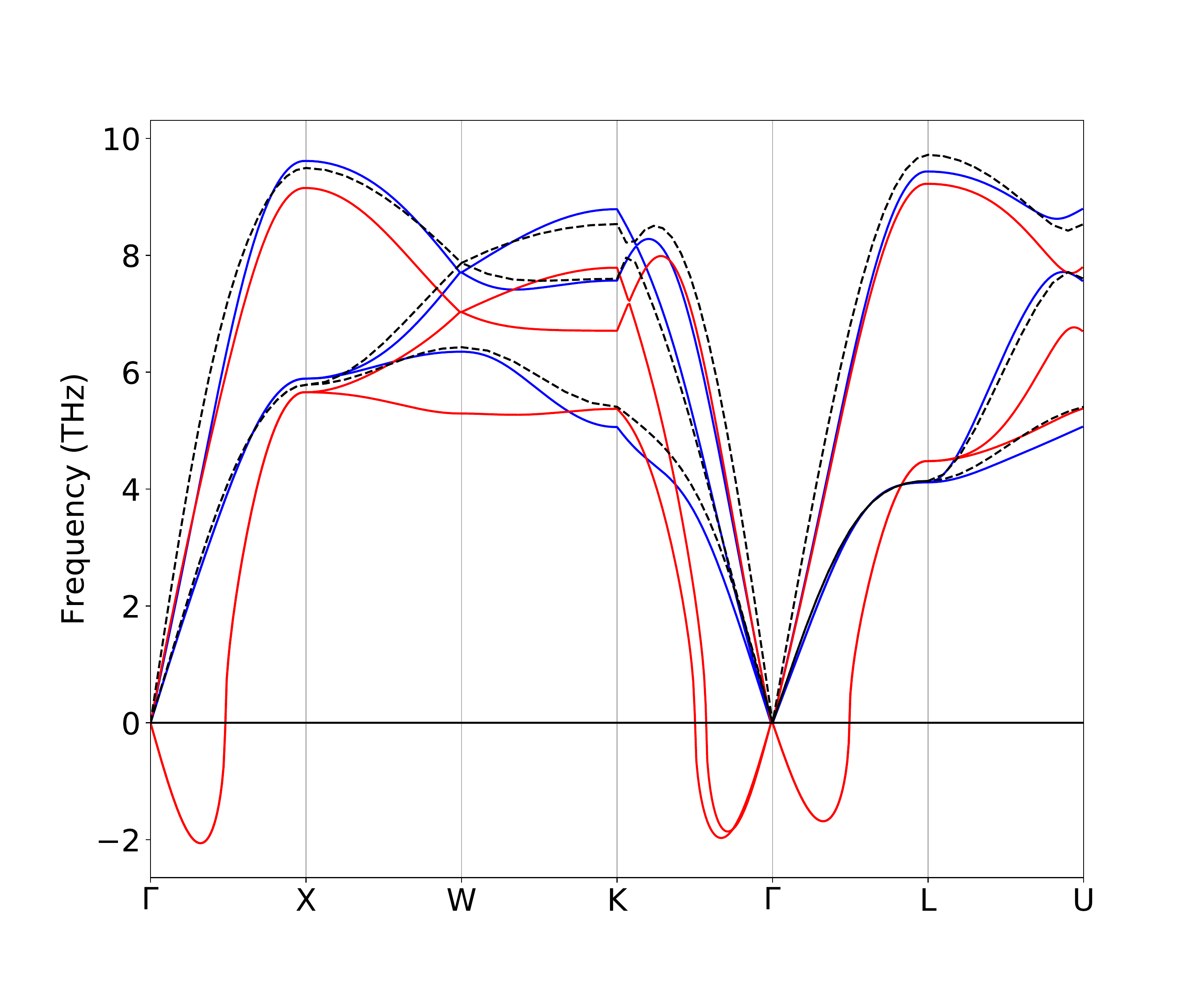}
    \caption{\acp{GAP} for Al based on \ac{NDSC} (blue line, 66.1 CPU hours) and diagonal supercell (red line, 60.1 CPU hours) data as compared against underlying \ac{DFT} (black dashed line) for a minimal amount of data.}
    \label{fig:ndsc_vs_diag_minimal}
\end{figure}

To illustrate the efficiency gains realised when using \ac{NDSC} configurations in the training set, we compare the phonon dispersion curves corresponding to two GAP models, constructed to emphasise the advantage of using configurations consisting of fewer atoms.
The two models were based on two separate databases, both of which required the same amount of computational time to calculate the \abinitio{} reference energies, forces and virial stresses.
To generate the first database, we used \ac{DSC} of size commensurate with the desired q-point sampling, where only a single atom is displaced from its ideal lattice site, as suggested by George \etal{}\cite{george_combining_2020}.
The other database contained \ac{NDSC} configurations generated using the workflow described in Section~\ref{sec:database}, such that the same amount of computational effort was needed to compute the \abinitio{} data as for the diagonal supercell.
The comparison of the phonon dispersion of the two models is presented in Figure~\ref{fig:ndsc_vs_diag_minimal}.


The model fitted on \ac{NDSC} configurations performs noticeably better, with phonon modes in close agreement with the \abinitio{} reference data. On the other hand, while the model fitted with diagonal supercells captures some of the phonon branches, it predicts unphysical dynamical instabilities in fcc Al. 

\begin{figure*}[htb]
    \centering
    \includegraphics[width=\linewidth]{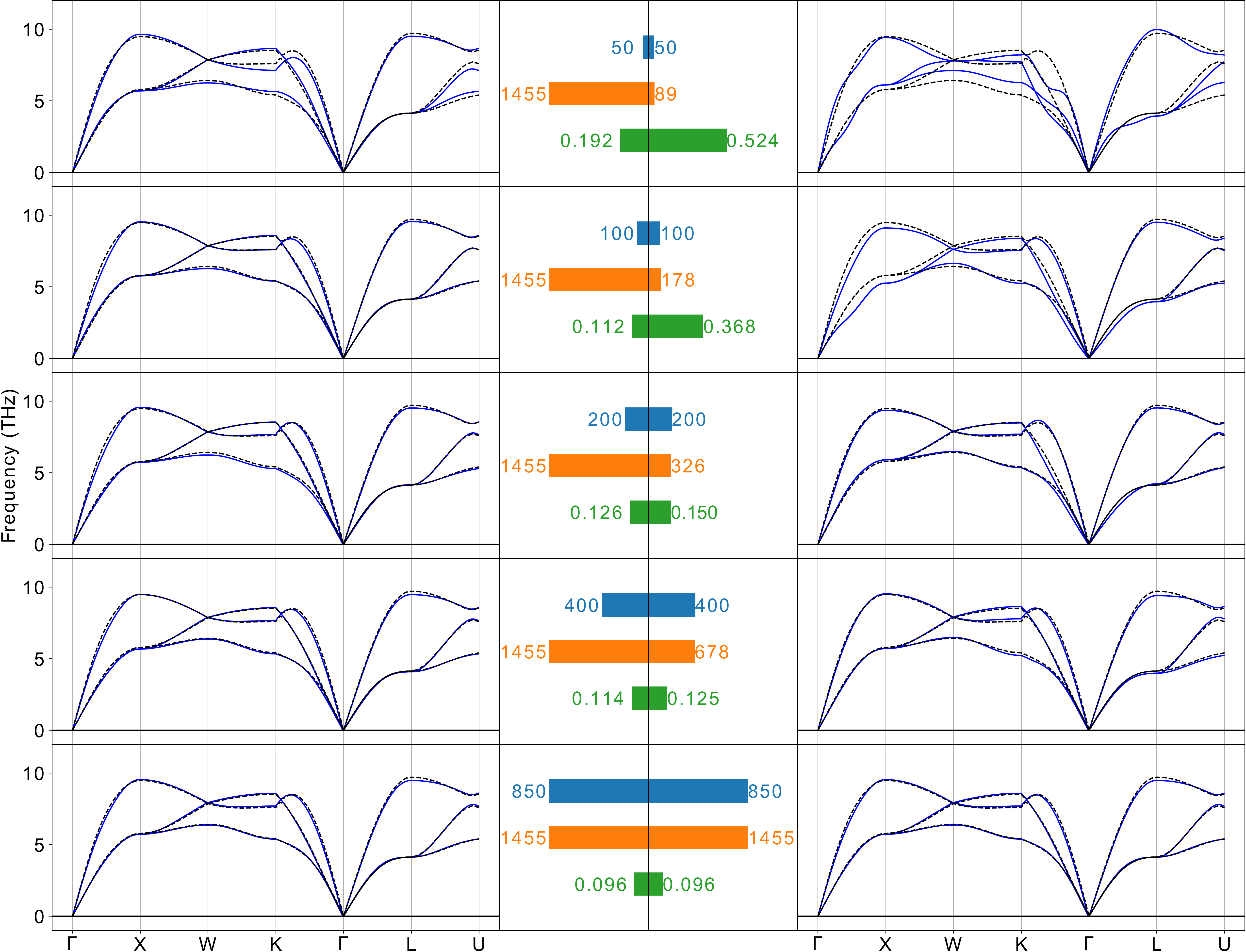}
    \caption{Left and right: Phonon dispersion curves of fcc Al \ac{GAP} models (blue) compared to the \ac{DFT} reference (black dashed).
    The set of training configurations increases from the top to the bottom, with the number of atomic environments displayed as blue bars (with numbers) on the centre panel. The orange bars represent the number of representative (sparse) points in each of the \ac{GAP} models, and the green bars show the \ac{RMSE} of the phonon frequencies in THz units.}
    \label{fig:fcc_Al_learning_curve}
\end{figure*}

To establish how the performance and accuracy of our \ac{MLIP} models benefit from increasing the amount of training data, we fitted a series of \ac{GAP} models for Al, using different size random subsets of the \ac{NDSC} data.
We present our learning curves as a series of phonon dispersion diagrams in Figure~\ref{fig:fcc_Al_learning_curve}, showing two approaches: 
\begin{enumerate*}[label=(\alph*)]
    \item keeping the set of representative atomic environments (or sparse points, set $M$ in equation~\ref{eq:total_energy}) constant across the models, using the representative set selected from atomic environments in the largest data set;
    \item selecting representative atomic environments from each of the fitting subset.
\end{enumerate*}
As expected, increasing the data size leads to significant but diminishing improvements the accuracy of the model, measured as RMSE of the predicted phonon dispersion against the \abinitio{} benchmark.
However, it is interesting to observe that when the sparse points, which represent basis functions in the \ac{GAP} framework, are selected from atomic environments not necessarily present in the training configurations, the accuracy is markedly improved even when using the same fitting targets.
Therefore we suggest that GAP models may be improved by adding atomic configurations that do not need \abinitio{} data associated with them, in order to increase the set of sparse points.
The advantage of this approach is that significantly less computational effort is needed to generate the expensive \abinitio{} data and it is possible to make improvements without the need to calculate additional target quantities at the \ac{DFT} level.



\begin{figure}[htb]
    \centering
    \includegraphics[width=\linewidth]{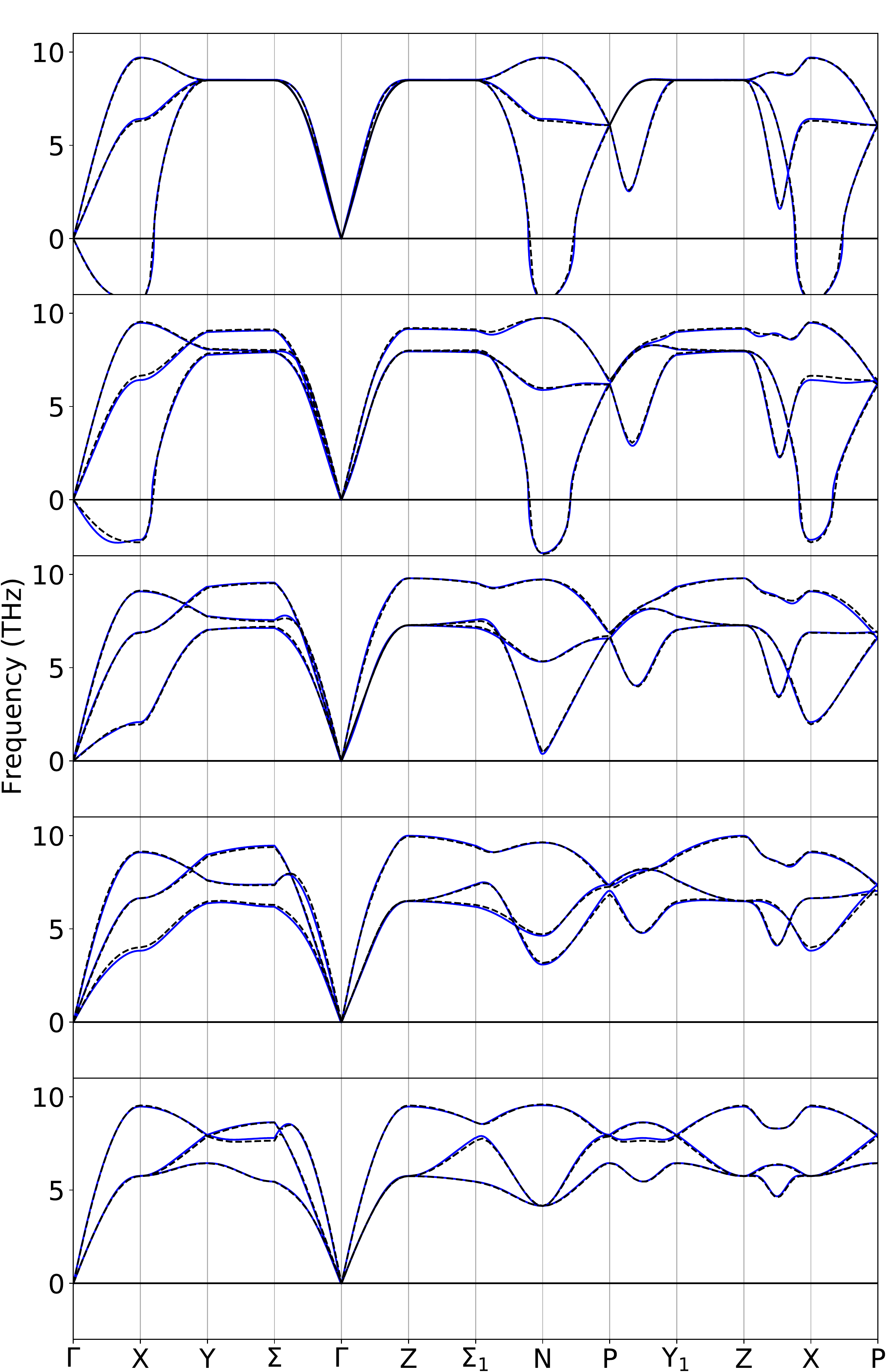}
    \caption{\ac{GAP} (blue) trained using \ac{NDSC} structures compared to commensurate \ac{DFT} (black dashed) phonon dispersion calculations for in Al, where the lattice was transformed from the stable fcc structure (bottom) to the dynamically unstable bcc structure (top) along the Bain path.}
    \label{fig:Al_bcc_phonons}
\end{figure}

We were also interested in studying how \ac{NDSC} configurations may assist fitting the \ac{PES} at stationary points other than minima.
Aluminium at ambient conditions is dynamically unstable in the body-centred cubic form, although at extreme pressures the bcc phase becomes energetically favourable \cite{sinrquotko_ab_2002}.
We have collected training configurations along the Bain path that connects the bcc and fcc phases of Al.
The lattice vectors of primitive unit cells were described as body-centred tetragonal,
\begin{equation}\label{eq:bainpath_mat}
\mathbf{L} =
\begin{bmatrix}
-\sfrac{a}{2} & \sfrac{a}{2} & \sfrac{c}{2} \\
\sfrac{a}{2} & -\sfrac{a}{2} & \sfrac{c}{2} \\
\sfrac{a}{2} & \sfrac{a}{2} & -\sfrac{c}{2} \\
\end{bmatrix}
\end{equation}
where the rows of $\mathbf{L}$ represent the cell vectors, and $a$ and $c$ change as
\begin{align*}
    a &= a_0 (1-\eta + 2^{\sfrac{1}{3}} \eta) \\
    c &= \frac{a_0}{\sqrt{1-\eta + 2^{\sfrac{1}{3}} \eta}}
\end{align*}
with $0 \le \eta \le 1$, such that the volume of the cell $|\det{\mathbf{L}}|$ is conserved.
With this definition, the cases $\eta=0$ and $\eta=1$ correspond to the bcc and fcc lattices, respectively.
We fitted a \ac{GAP} model based on perturbed \ac{NDSC} configurations that were generated using primitive unit cells at $\eta=0, \sfrac{1}{2} \textrm{ and } 1$. 
We benchmarked the phonon dispersion curves obtained with our model considering intermediate $\eta$ values, as shown in Figure~\ref{fig:Al_bcc_phonons}.
We found excellent agreement with \ac{DFT} overall, with instabilities at the $N$-point reproduced highly accurately.

\begin{figure}[htb]
    \centering
    \includegraphics[width=\linewidth]{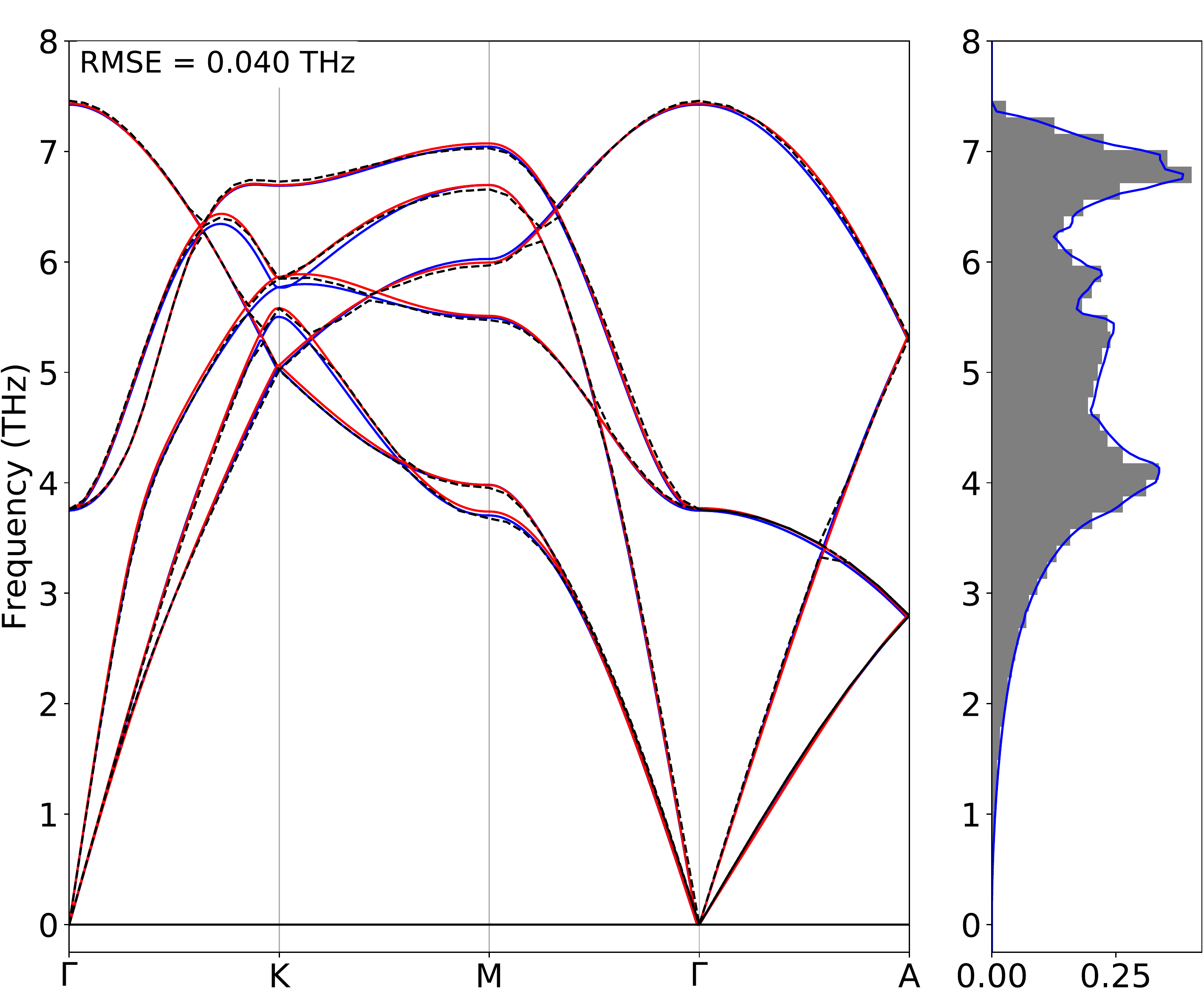}
    \caption{\acp{GAP} trained using \ac{NDSC} (blue) with component wise regularisation as informed by the force on each atom and virial stress on each configuration compared to a model where the regularisation is set to a constant value (red line) as well as \ac{DFT} (black dashed) phonon dispersion curves.}
    \label{fig:Mg_hcp_phonons}
\end{figure}

We have also included a non-cubic crystalline system, magnesium, in our benchmarks, whose ground state structure at ambient condition is hexagonal close packed.
Using \ac{NDSC} configurations to train a \ac{GAP} model, we find excellent agreement between the \ac{GAP} and the \abinitio{} phonon dispersion curves.

\begin{figure}[htb]
    \centering
    \includegraphics[width=0.85\linewidth]{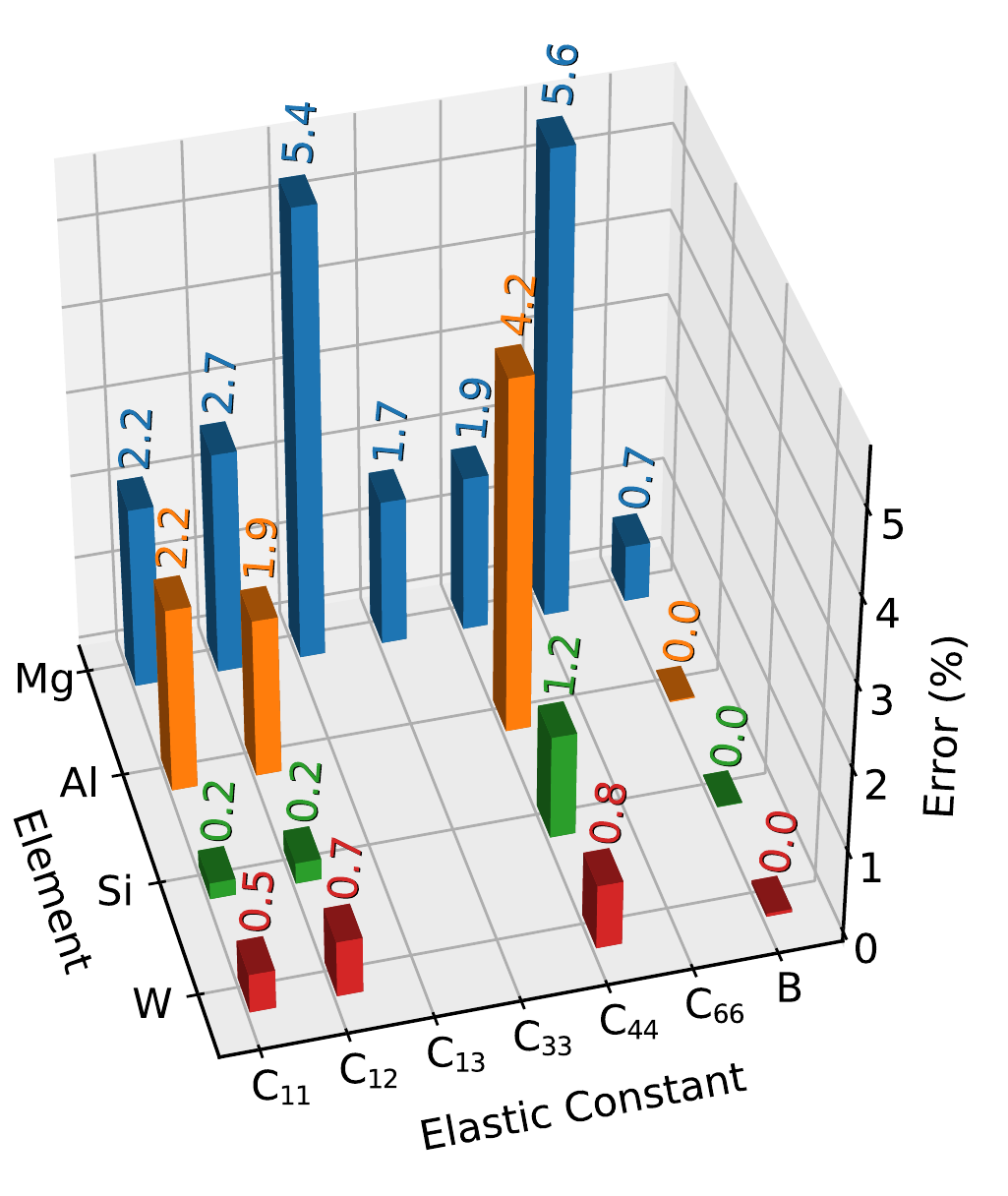}
    \caption{Relative error of elastic constants of \ac{GAP} models of Mg (blue), Al (orange), Si (green) and W (red) compared to \ac{DFT} results.}
    \label{fig:elastic_constant_err}
\end{figure}

While accurate phonon spectra on acoustic modes near the $\Gamma$-point indicate that the elastic properties of the crystal are well represented\cite{Boer.2018}, it is insightful to directly examine the numerical values of the elastic constants of the \ac{MLIP} models for additional benchmarking purposes. Table~\ref{tab:geom_elastic} and Figure~\ref{fig:elastic_constant_err} summarise the elastic moduli computed at the relaxed geometries both using \ac{DFT} and the \ac{GAP} models, showing excellent agreement with up to 6\% error.

\begin{figure}[htb]
    \centering
    \includegraphics[width=0.85\linewidth]{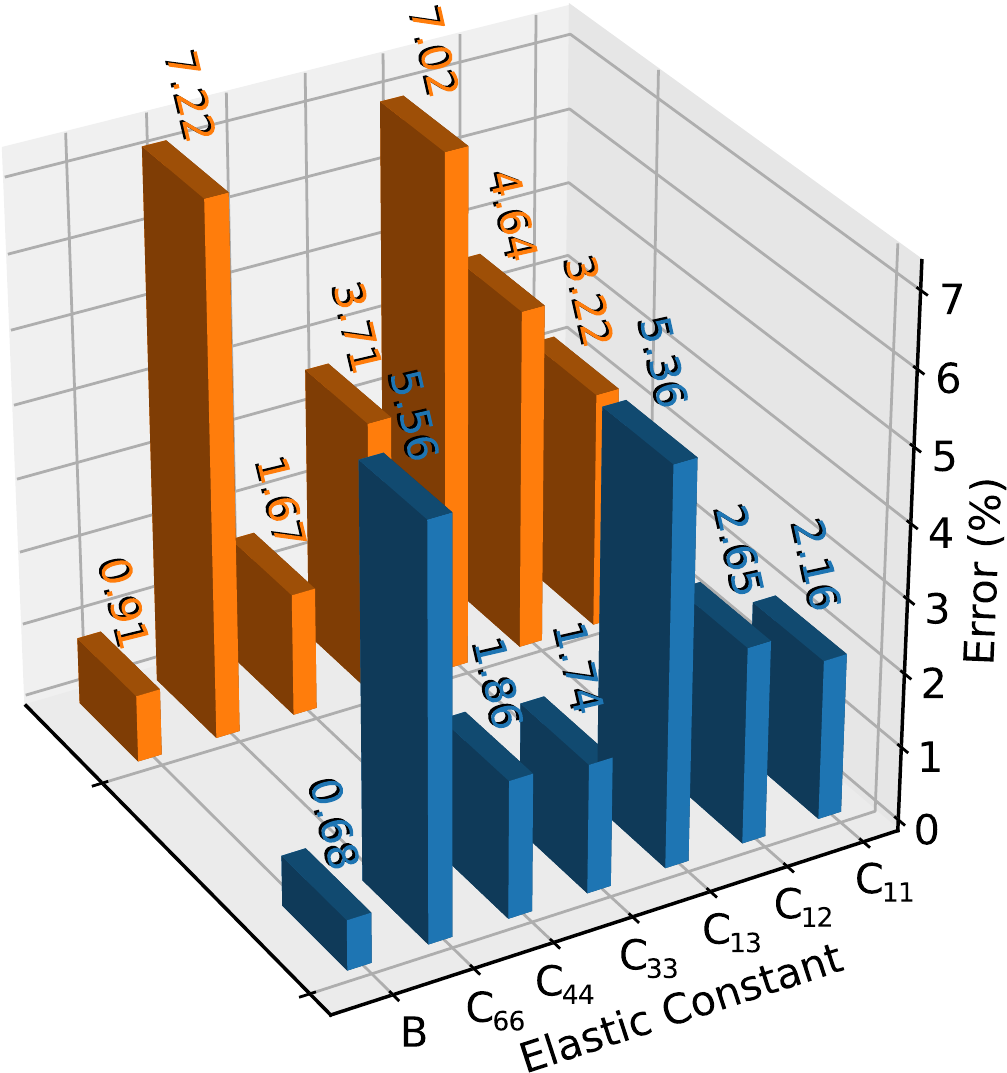}
    \caption{Errors of elastic constants of \ac{GAP} models of Mg relative to \ac{DFT} values in the hcp crystal structure. \ac{GAP} models were fitted using static force and virial regularisation (orange bars and figures) and adaptive force and virial regularisation (blue bars and figures), as described in section~\ref{sec:fitting}.
    }
    \label{fig:Mg_elastic_constant_err_reg}
\end{figure}

We illustrate the effect of adaptive regularisation of individual virial stress components, introduced in Section~\ref{sec:database}, by comparing two Mg \ac{GAP} models, one of which uses a static regularisation of 0.01~eV for each virial component, while the other employing the adaptive scheme.
As shown in Figure~\ref{fig:Mg_elastic_constant_err_reg}, some of the elastic constants are only reproduced to an error of up to 7\%, while introducing the adaptive virial regularisation, accuracy is significantly improved across all elastic constants without any deterioration of the quality of the phonon dispersion curves depicted in Figure~\ref{fig:Mg_hcp_phonons}.

\section{Conclusions}
In conclusion, we explored a computationally efficient approach using the \ac{NDSC} method introduced by Lloyd-Williams and Monserrat to generate database configurations for fitting \ac{MLIP} models based on \abinitio{} data.
We found that \ac{NDSC} configurations provide sufficient data to fit \acp{MLIP} reproducing the \ac{FCM} near stationary points of bulk crystalline materials, while costing significantly less computational effort than diagonal supercells.
We have also suggested an adaptive scheme to regularise virial stress components of \abinitio{} databases and demonstrated improvements of the elastic behaviour of \acp{MLIP}.
The procedure described in this work can be fully automated and integrated into existing database generating workflows, allowing to save computational cost or include a greater variety of representative structures, realising savings on cost and carbon emissions associated with high-performance computations, or improved quality \acp{MLIP}.
We also envisage further use of \ac{NDSC} configurations in databases used to inform models for alloy materials, as an addition or alternative to semi quasi-random structures, where the substitution of elemental species may be regarded as alchemical perturbations.

\section{Software and data availability}
All \abinitio{} training data and the scripts used to generate the configurations are made available in a dedicated repository\cite{repo}.
We used the QUIP software package with the \ac{GAP} plugin\cite{quip}, available under the General Public License and the Academic Source License, respectively. The Atomic Simulation Environment\cite{Larsen.2017} was used to manipulate atomic configurations and we employed the \verb+phonopy+ package\cite{phonopy} to calculate the phonon dispersion curves of the \ac{GAP} models.
Our workflow greatly benefited from using GNU \verb+parallel+\cite{parallel}.

\section{Acknowledgments}
We thank Noam Bernstein and Keith Refson for fruitful discussions.
We acknowledge support from the NOMAD Centre of Excellence, funded by the European Commission under grant agreement 951786.
CA is supported by a studentship jointly by the UK Engineering and Physical Sciences Research Council–supported Centre for Doctoral Training in Modelling of Heterogeneous Systems, Grant No. EP/S022848/1 and the Atomic Weapons Establishment.
ABP acknowledges funding from CASTEP-USER funded by UK Research and Innovation under the grant agreement EP/W030438/1.
Calculations were performed using the Sulis Tier 2 HPC platform hosted by the Scientific Computing Research Technology Platform at the University of Warwick. Sulis is funded by EPSRC Grant EP/T022108/1 and the HPC Midlands+ consortium.
We acknowledge the University of Warwick Scientific Computing Research Technology Platform for assisting the research described within this study.

\bibliography{bib.bib}
\bibliographystyle{apsrev4-1}

\end{document}